\documentclass[10pt]{article}
\usepackage[OME]{express}

\begin{document}

\title{Toward theoretically limited SPP propagation length above two hundred microns on ultra-smooth silver surface}

\author{Aleksandr S. Baburin,\authormark{1,2,7,*} Aleksey S. Kalmykov,\authormark{3,4,*} Roman V. Kirtaev,\authormark{5} Dmitriy V. Negrov,\authormark{5} Dmitriy O. Moskalev,\authormark{1} Ilya A. Ryzhikov,\authormark{1,6} Pavel N. Melentiev,\authormark{3,4,8} Ilya A. Rodionov,\authormark{1,2} and Victor I. Balykin\authormark{3,4,9}}

\address{\authormark{1}Functional Micro/Nanosystems Research and Educational Center, Bauman Moscow State Technical University, 2nd Baumanskaya steet 5, Moscow, 105005, Russian Federation\\
\authormark{2}Dukhov Research Institute of Automatics (VNIIA), Sushchevskaya street 22,Moscow 127055, Russian Federation\\
\authormark{3}Institute of Spectroscopy RAS, Troitsk, Moscow, 142190, Russia\\
\authormark{4}National Research University, Higher School of Economics, Moscow, 101000, Russia\\
\authormark{5}Moscow Institute of Physics and Technology, Dolgoprudny, 141700, Russia\\
\authormark{6}Institute for Theoretical and Applied Electromagnetics RAS, Izhorskaya street 13, Moscow, 125412, Russian Federation\\}
\email{\authormark{7}baburin@bmstu.ru} 
\email{\authormark{8}melentiev@isan.troitsk.ru}
\email{\authormark{9}balykin@isan.troitsk.ru}
\address{\authormark{*}A.S. Baburin and A.S. Kalmykov are contributed equally to this work.}



\begin{abstract}
We demonstrate the optical medium for surface plasmon - polariton waves (SPP) propagation with ultra low losses corresponding to the theoretically limited values. The unique element of the optical medium is an atomically-flat single-crystalline silver thin film which provides extremely low losses. The SPP excited on the surface of such thin films ($\lambda$ = 780 nm) is characterized by a SPP propagation length equal to 200 $\mu$m, which is twice longer than previously reported experimental results and corresponds to theoretically limited values for silver films.\end{abstract}


1. J. B. Khurgin, "How to deal with the loss in plasmonics and metamaterials," Nat. nanotechnology 10, 2 (2015).

2. V. I. Balykin and P. N. Melentiev, "Optics and spectroscopy of individual plasmonic nanostructure," Phys. Usp (2017).

3. P. Melentiev, A. Kalmykov, A. Gritchenko, A. Afanasiev, V. Balykin, A. Baburin, E. Ryzhova, I. Filippov, I. Rodionov, I. Nechepurenko, A. Dorofeenko, I. Ryzhikov, A. Vinogradov, A. Zyablovsky, E. Andrianov, and A. Lisyansky, "Plasmonic nanolaser for intracavity spectroscopy and sensorics," Appl. Phys. Lett. 111, 213104 (2017).

4. J. Khurgin and G. Sun, "Third-order nonlinear plasmonic materials: Enhancement and limitations," Phys. Rev. A 88, 053838 (2013).

5. I. Radko, S. I. Bozhevolnyi, G. Brucoli, L. Martin-Moreno, F. Garcia-Vidal, and A. Boltasseva, "Efficient unidirectional ridge excitation of surface plasmons," Opt. express 17, 7228-7232 (2009).

6. P. Melentiev, A. Kuzin, D. Negrov, and V. Balykin, "Diffraction-limited focusing of plasmonic wave by a parabolic mirror," Plasmonics pp. 1-7 (2018).

7. T. Holmgaard and S. I. Bozhevolnyi, "Theoretical analysis of dielectric-loaded surface plasmon-polariton waveguides," Phys. Rev. B 75, 245405 (2007).

8. P. Melentiev, A. Kuzin, V. Balykin, A. Ignatov, and A. Merzlikin, "Dielectric-loaded plasmonic waveguide in the visible spectral range," Laser Phys. Lett. 14, 126201 (2017).

9. M. Noginov, G. Zhu, A. Belgrave, R. Bakker, V. Shalaev, E. Narimanov, S. Stout, E. Herz, T. Suteewong, and
U. Wiesner, "Demonstration of a spaser-based nanolaser," Nature. 460, 1110 (2009).

10. S. Bogdanov, M. Y. Shalaginov, A. Lagutchev, C.-C. Chiang, D. Shah, A. Baburin, I. Ryzhikov, I. Rodionov, A. V. Kildishev, A. Boltasseva, and V. M. Shalaev, "Ultrabright room-temperature sub-nanosecond emission from single nitrogen-vacancy centers coupled to nano-patch antennas," Nano letters (2018).

11. S. I. Bozhevolnyi and J. B. Khurgin, "The case for quantum plasmonics," Nat. Photonics 11, 398 (2017).

12. G. Yankovskii, A. Komarov, R. Puzko, A. Baryshev, K. Afanasiev, I. Boginskaya, I. Bykov, A. Merzlikin, I. Rodionov, and I. Ryzhikov, "Structural and optical properties of single and bilayer silver and gold films," Phys. Solid State 58, 2503-2510 (2016).

13. K. M. McPeak, S. V. Jayanti, S. J. Kress, S. Meyer, S. Iotti, A. Rossinelli, and D. J. Norris, "Plasmonic films can easily be better: rules and recipes," ACS photonics 2, 326-333 (2015).

14. S. Kunwar, M. Sui, Q. Zhang, P. Pandey, M.-Y. Li, and J. Lee, "Various silver nanostructures on sapphire using plasmon self-assembly and dewetting of thin films," Nano-Micro Lett. 9, 17 (2017).

15. A. S. Baburin, A. I. Ivanov, I. A. Ryzhikov, I. V. Trofimov, A. R. Gabidullin, D. O. Moskalev, Y. V. Panfilov, and I. A. Rodionov, "Crystalline structure dependence on optical properties of silver thin film over time," in "Progress In Electromagnetics Research Symposium-Spring (PIERS), 2017," (IEEE, 2017), pp. 1497-1502.

16. I. A. Rodionov, A. S. Baburin, A. P. Vinogradov, A. R. Gabidullin, S. S. Maklakov, S. Peters, I. A. Ryzhikov, and A. V. Andriyash, "Synthesis of low-loss single-crystalline silver films for quantum nanophotonics," arXiv preprint arXiv:1806.07611 (2018).

17. P. R. West, S. Ishii, G. V. Naik, N. K. Emani, V. M. Shalaev, and A. Boltasseva, "Searching for better plasmonic materials," Laser and Photonics Rev. 4, 795-808 (2010).

18. J.-S. G. Bouillard, W. Dickson, D. P. O Connor, G. A. Wurtz, and A. V. Zayats, "Low-temperature plasmonics of metallic nanostructures," Nano letters 12, 1561-1565 (2012).

19. S. V. Jayanti, J. H. Park, A. Dejneka, D. Chvostova, K. M. McPeak, X. Chen, S.-H. Oh, and D. J. Norris,
"Low-temperature enhancement of plasmonic performance in silver films," Opt. materials express 5, 1147-1155
(2015).

20. S. A. Maier, Plasmonics: fundamentals and applications (Springer Science and Business Media, 2007).

21. J. Sukham, O. Takayama, A. V. Lavrinenko, and R. Malureanu, "High-quality ultrathin gold layers with an aptms adhesion for optimal performance of surface plasmon polariton-based devices," ACS applied materials and interfaces 9, 25049-25056 (2017).

22. H. Raether, "Surface plasmons on smooth surfaces," in “Surface plasmons on smooth and rough surfaces and on
gratings," (Springer, 1988), pp. 4-39.

23. P. B. Johnson and R.-W. Christy, "Optical constants of the noble metals," Phys. review B 6, 4370 (1972).

24. Y. Wu, C. Zhang, N. M. Estakhri, Y. Zhao, J. Kim, M. Zhang, X.-X. Liu, G. K. Pribil, A. Alu, C.-K. Shih, and X. Li, "Intrinsic optical properties and enhanced plasmonic response of epitaxial silver," Adv. Mater. 26, 6106-6110 (2014).

25. S. Babar and J. Weaver, "Optical constants of Cu, Ag, and Au revisited," Appl. Opt. 54, 477–481 (2015).

26. P. Melentiev, A. A. Kuzin, and V. I. Balykin, "Control of SPP propagation and focusing through scattering from nanostructures," Quantum Electron. 47, 266 (2017).

27. I. A. Rodionov, A. S. Baburin, A. V. Zverev, I. A. Philippov, A. R. Gabidulin, A. A. Dobronosova, E. V. Ryzhova, A. P. Vinogradov, A. I. Ivanov, S. S. Maklakov, A. Baryshev, I. Trofimov, A. Merzlikin, N. Orlikovsky, and I. Rizhikov, "Mass production compatible fabrication techniques of single-crystalline silver metamaterials and plasmonics devices," in "Metamaterials, Metadevices, and Metasystems 2017," , vol. 10343 (International Society for Optics and Photonics, 2017), vol. 10343, p. 1034337.

28. F. Lopez-Tejeira, S. G. Rodrigo, L. Martin-Moreno, F. J. Garcia-Vidal, E. Devaux, T. W. Ebbesen, J. R. Krenn, I. Radko, S. I. Bozhevolnyi, M. U. Gonzalez, J. Weeber, and A. Dereux, "Efficient unidirectional nanoslit couplers for surface plasmons," Nat. Phys. 3, 324 (2007).

\section{Introduction}
One of the key performance criteria in nanoplasmonics is the plasmon waves propagation on the surface of metal thin films. Most nanoplasmonic practical applications are strongly limited by dramatic losses of SPP waves on metal surfaces [1-3]. The losses in metals inevitably limit the propagation length of the SPP wave, the Q factor of plasmonic resonators, and the cross section of optical nonlinear processes [4]. This problem is even worse in nanoplasmonics of the visible and near infrared spectral ranges. For example, a short propagation length of SPP leads to low efficiency of SPP focusing [5, 6], waveguiding [7, 8], higher thresholds of plasmonic nanolasers [3, 9], and lower Purcell factor of quantum nanoplasmonics resonators [10, 11].

There are many experimental researches demonstrating a strong difference for various metal films between measured SPP losses and calculated values even for well-known or measured optical constants [1, 2]. Such a huge mismatch is a result of the metal thin films (crystalline structure, surface roughness, surface condition) and interfaces (substrate-metal, metal-media) imperfections. Single-crystalline metal thin films should be a proper way to solve described problems, but it is extremely hard to fabricate them with high quality. One of the most challenging metal in terms of single-crystalline growth is silver, because of its natural chemical instability [12, 13], high sensitivity to substrate surface reconstruction and impurities, lattice-matched substrate dewetting at elevated growth temperatures [14] and optical properties degradation over time [15]. All the listed above silver film growth issues are in practice the potential sources of films poor optical properties, characterized by a short SPP propagation length.

In this work we demonstrate for the first time that using of recently developed method of atomically-flat single-crystalline silver thin film growth [16] is a proper solution to create the optical medium for SPP propagation with losses close to theoretically predicted values. Such a single-crystalline silver film can be obtained utilizing the high vacuum physical vapour deposition two-step growth approach named as the SCULL process (Single-crystalline Continuous Ultra-flat Low-loss Low-cost). The fundamental idea of the process involves a growth of an effective underlying layer under elevated temperature (first step), which emulates a lattice-matched substrate for the targeted metal, followed by a single-crystalline film growth under room temperature (second step) on the just synthesized lattice-matched substrate in the same vacuum cycle. The process provides the single-crystalline metallic films growth on non-ideally lattice-matched substrates without underlayers using a standard high vacuum electron-beam evaporator.

\section{SPP propagation length on the silver film surface}
The material choice for creating plasmonic nanostructures is decisive in achieving its best properties. Silver and gold at the present time are the main materials of experimental nanoplasmonics due to minimal ohmic losses among all known natural materials. However, these materials in visible and ultraviolet regions of spectrum exhibit significant ohmic losses. This situation eliminates the possibility to construct high performance useful nanoplasmonic devices in these spectral ranges. 

The ohmic losses in metals can be divided into two groups: (1) losses due to the presence of free conduction electrons and (2) losses due to bound electrons in metal. Losses from conduction electrons have different origins: electron-electron interaction, electron-phonon interaction, and due to scattering of electrons on lattice defects and grain boundaries in polycrystalline metals. Losses due to bound electrons arise when the photon is absorbed through the interband transition, with the electron transition to highly excited states. Both mechanisms of losses in metals strongly limit the progress in the development of nanoplasmonics.

Using silver one could obtain the minimal losses for SPP waves in the visible and near infrared regions of spectrum [17]. A further reduction in losses can be achieved by cooling the material to cryogenic temperatures [18] with about 40-60\% improvement in the SPP propagation length [19].  These results are consistent with the Drude-Lorentz model of the metal.

The SPP propagation length is a key parameter characterizing plasmonic quality of metal films for future practical applications [20]. In this work we do measurements of the SPP propagation length to characterize plasmonic properties of Ag film under study. Let's pay attention to another approach sensitive to imperfections of metal films based on measurements of short range SPPs dispersion diagram [21]. The SPP propagation length is the distance where SPP wave intensity decreases to 1/e from its initial value [22]. The theoretical value of the SPP propagation length is determined by the optical constants of silver ($\varepsilon^{'}$ - the real part, and $\varepsilon^{''}$ - the imaginary part of the dielectric permittivity of silver film):
\begin{equation}
L_{SPP} = \frac{c}{\omegaup}\left (\frac{\varepsilon^{'}+1}{\varepsilon^{'}}\right)^{3/2}\frac{(\varepsilon^{'})^2}{\varepsilon^{''}}
\end{equation}

Thus, to calculate the SPP propagation length, it is necessary to know the dielectric constants of silver film. There are numerous studies  on silver films optical constants measured using spectroscopic ellipsometry. In the work [23], the optical constants were measured at wavelength of 780 nm that are $\varepsilon^{'}$= -29.384, $\varepsilon^{''}$ = 0.365222.
The corresponding calculated SPP propagation length is 278 $\mu$m ($\lambda$ = 780 nm). Another value of the SPP propagation length can be obtained by using the following data from [24]: $\varepsilon^{'}$= -31.535, $\varepsilon^{''}$ = 0.4328. 
The corresponding calculated SPP propagation length is 271 $\mu$m ($\lambda$ = 780 nm). The minimum calculated value of the SPP propagation length is obtained using the data of [25]: $\varepsilon^{'}$= -29.688, $\varepsilon^{''}$ = 0.57674, which corresponds to the SPP propagation length equals to 180 $\mu$m ($\lambda$ = 780 nm). However, till the present time,  the largest experimentally measured SPP propagation length for silver films is about 100 $\mu$m ($\lambda$ = 780 nm) [26], which is several times less than the calculated values.

In all the experiments we used 100-nm-thick single-crystalline silver films [16] grown on a lattice-matched Si [111] substrates with atomically flat surface (typical AFM root mean square roughness is less than 0.5 nm measured over 2.5 $\times$ 2.5 $\mu$m area). The films were deposited by electron-beam evaporation (base pressure $3 \times 10^{-8}$ Torr, 1 $\Dot{A}/s$ deposition rate) using 99.999\% silver pellets. Using AFM and XRD measurements we confirm the evidence of atomically flat single - crystalline properties of the film, similar to published in [16]. The silver films demonstrate perfect optical properties combined with nanostructures patterning compatibility [10, 27].In this paper we present the SPP propagation length measurement results for such a single-crystalline silver thin film.

\section{Measurement of the SPP propagation length on the surface of single-crystalline grown silver film}
Measurement of the SPP propagation length is a rather complicated experimental problem and usually consists of two steps: (1) effective excitation of SPP wave and (2) measurement of the SPP propagation length. One of the most accessible, convincing and informative is the method of measuring the SPP propagation length based on registration of SPP waves which are excited by laser radiation. The SPP are excited by using a nanoslits array patterned on a metal film surface. Registration of the launched SPP is performed through the scattering the SPP on the nanogrooves (detectors of SPP waves) located on the metal film surface on the SPP propagation way Fig.1(c) [26]. 

We call this method "the far-field optical microscopy of SPP waves". In this method, the excitation of SPP is realized through the scattering of laser radiation on the array of nanoslits and the subsequent scattering of the launched SPP on nanogrooves. Detection of scattered radiation on nanogrooves in far field helps to visualize SPP propagation as well as to measure the SPP propagation length. 

The measurements were made with silver films both on transparent [26], and on opaque substrates [24]. We note that optical microscopy of SPP waves on metallic films made on an opaque substrate is much more complicated in implementation, in comparison with the case of transparent substrates. This complexity is caused by the fact that excitation of the SPP and its detection is carried out from the same side of the substrate. In this case, the same microscope objective is used to excite SPP and then to detect radiation scattered by the SPP on nanostructures - detectors. The main problem here is to detect a weak scattering signal of SPP on the nanostructures - detectors against a giant reflected and scattered signals from the laser used to excite SPP.

Two types of nanostructures were created on the silver film surface using focused-ion-beam milling. One of them was designed to excite SPP wave and was formed by a matrix of 20 slits nanostructures with a nanoslits spacing of $\Lambda_1$= 780 nm and a slit size of 120 nm $\times$ 20 $\mu$m. Another one served as a detector of the SPP wave, and was formed by a matrix of 15 nanogrooves with a 40 nm depth, 120 nm width and 40 $\mu$m length with a distance between nanogrooves equal to 20 $\mu$m. Fig. 1(a,b) shows the electron microscope images of the created nanostructures.

\begin{figure}[ht!]
\centering\includegraphics[width=11cm]{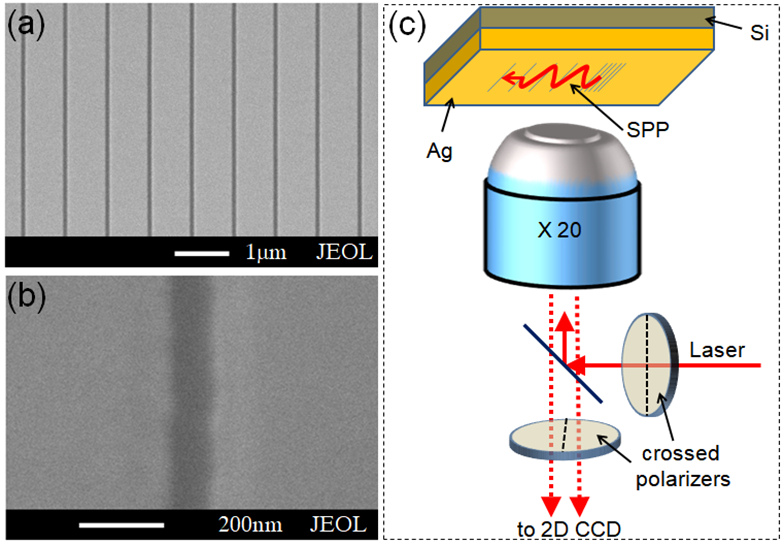}
\caption{(a) an electron microscope image of nanoslits array on the Ag film used to excite SPP, (b) an electron microscope image of a nanogroove used to detect SPP, (c) a schematic diagram of the experimental set up.}
\end{figure}

Fig.1(c) shows the scheme for measuring the SPP propagation length. The measurements were performed using an inverted Nikon Eclipse/Ti-U microscope. To excite SPP wave, we used a CW semiconductor laser with a tunable wavelength around 780 nm, having ultra narrow linewidth. This laser light was focused with a microscope objective (x20) into a 6 $\mu$m spot on the array of nanoslits. The period of nanoslits array $\Lambda_1$ was chosen to effectively excite SPP on the film surface at orthogonal incidence of the laser beam on the sample [28]. The period was determined from the relation $Re(k_{SPP})$ $\thickapprox$ $G + k_0\sin\alpha$, where $k_{SPP}$ is the wave number of the SPP wave, G = $2\pi/\Lambda_1$ is the modulus of the reciprocal lattice vector of the nanoslits array, $\alpha$  is the angle of laser radiation incidence on the sample ($\alpha=90^0$), $k_0 =  \omega c$  is the wavenumber of laser light.

We fabricated series of parallel nanogrooves on the way of SPP propagation Fig.1(a). The radiation scattered on these nanogrooves is recorded with the same microscope objective lens on a 2D CCD camera (Princeton Instrument, PhotonMax). The radiation is proportional to the intensity of the SPP wave at the location of the nanogroove, and the detection of the scattered light from the array of nanogrooves allows us to measure a change of the SPP intensity during its propagation. To increase the contrast of the measured signal, we used the fact that the radiation formed due to scattering of SPP on nanogrooves has different polarization compared to incident laser light. We used two crossed polarizers having a residual extinction ratio of 1:1000. One of them was installed in the laser beam. Another one was installed just before CCD camera in such a way to block most of laser light reflected from the sample. Such arrangement of the polarizers helped us to substantially reduce laser radiation reflected and scattered from the sample.

Fig.2(a) shows the Ag film surface optical image with fabricated nanostructures	which were fabricated using focus-ion-beam milling method. The figure clearly shows both types of nanostructures. One is the array of nanoslits, designed to excite the SPP and which is visible as a rectangle. The distance between nanoslits is too small to be resolved by optical microscope. The figure also clearly shows distinctly separated nanogrooves. These nanogrooves are arranged parallel to the nanoslits and are on the way of the SPP propagation. Scattering of the SPP on each nanogroove allows us to visualize the SPP, and also to measure the SPP propagation length.

\begin{figure}[ht!]
\centering\includegraphics[width=11cm]{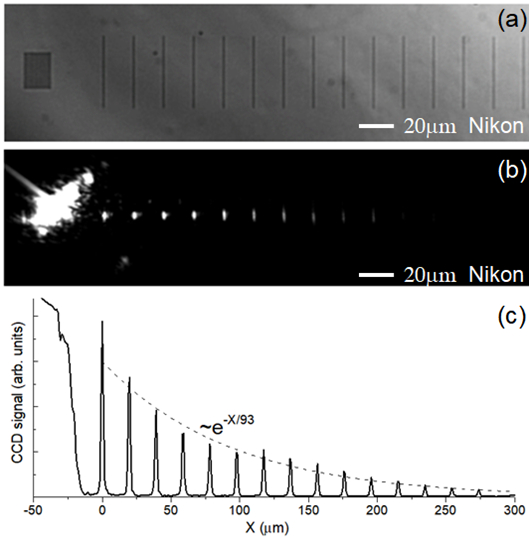}
\caption{Optical microscopy of SPP: (a) optical image of the Ag film surface with nanostructures created by a focused-ion-beam lithography, (b) optical image of the Ag film surface when SPP is excited by a laser radiation, (c) the cut image of Fig.2(b) along the SPP propagation direction; the dotted curve is the approximation by exponential curve with the characteristic decay length of 93 $\mu$m.}
\end{figure}

Figure 2b shows the optical image of the silver film surface when it is irradiated by laser radiation at wavelength 780 nm. As it can be seen from the figure, the strong scattering of laser radiation on the array of nanoslits and the excitation of SPP on the Ag film surface appear. The SPP propagates and scatters on nanogrooves, giving possibility to visualize SPP propagation in optical microscope. Each of the spot on a nanogroove has elliptical shape. The smaller diameter $d_1$ of this spot is determined by a diffraction limit of used optical objective. The larger diameter of the spot $d_2$ is determined by width of SPP. Thus the optical image clearly shows the divergence of the SPP due to diffraction: as the distance from the excitation region of the SPP wave increases, the spot diameter $d_2$ (from the scattering of the SPP wave by the nanogrooves) becomes higher. The spot size on a nanogroove located close to the array of nanoslit is $d_2$(0 $\mu$m) = 6 $\mu$m and corresponds to the diameter of the exciting laser beam. On the nanogroove located at L = 275 $\mu$m from the first nanogroove, the corresponding spot size is much lager: $d_2$(275 $\mu$m) = 38 $\mu$m.

Fig.2(c) shows a cut image of Fig.2(b) along the SPP propagation direction. As can be seen from the Fig.2(b), the scattering signals from the nanogrooves are represented by the narrow peaks with the width equal to $d_1$ = 2 $\mu$m, which is determined by the resolution of the objective lens with a numerical aperture NA = 0.4 ($d_1\thickapprox\lambda/NA$).

It can be seen from Fig.2(b,c) that the amplitudes of the scattering signal of SPP on the nangrooves are different and decrease with a distance between each nanogroove and the array of nanoslits because of the losses of the SPP in Ag thin film. The change in the amplitudes is well approximated by an exponential curve with a characteristic length equal to 93 $\mu$m. This measured SPP decay length is determined by three factors: (1) absorption losses of the SPP in the silver film, (2) losses of the SPP due to scattering on each nanogroove, and (3) divergence of the SPP wave caused by its diffraction. We will show below that from these measurements it is possible to extract the SPP propagation length determined only by the losses in the silver film. 

\begin{figure}[ht!]
\centering\includegraphics[width=11cm]{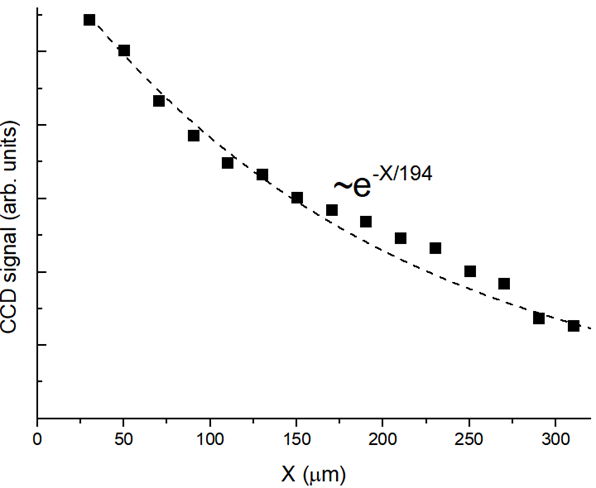}
\caption{The decay curve of the SPP on an Ag thin film surface.}
\end{figure}

Using the method described in [26], we measured losses of a SPP on each nanogroove, which equals to 4.8\%. Briefly, to measure the losses we have created another set of structures with a distance between nanogrooves two times lower compared to the structure presented  on the Fig. 2. Thus the SPP crosses two times more nanogrooves during its propagation compared to the case presented on Fig. 2. The set of nanoslits and nanogrooves was created on the same Ag film with a distance about 0.5 mm from the structures presented on Fig. 5. For the measurements it is important that the nanogrooves are identical in the both structures. This was controlled by electron microscopy. To measure losses of SPP on nanogrooves we measured SPP decay curve Fig. 2(c) for the both structures. The measured difference was attributed to losses of SPP on nanogrooves because number of nanogrooves is different and it permits to find losses of SPP on individual nanogroove [26].

The SPP diffraction was taken into account by integrating the signal along each nanogroove, using the image of Fig.2(b), which corresponds to scattering of the SPP by the nanogroove. Fig.3 shows the dependence of the SPP intensity attenuation, taking into account the losses at each nanogroove and the diffraction of the SPP wave. As can be seen from the figure, the data are well approximated by an exponential curve with an attenuation length characterizing SPP propagation length on a surface of Ag film $L_{SPP}$ = 194 $\pm$ 23 $\mu$m. In this figure, the zero position on the x axis corresponds to the edge of the nanoslits array (used to excite the SPP). It is also seen from the figure that even at a distance of 300 $\mu$m from the excitation region, the SPP still contains a significant amount of energy to be easily detected.

\section{Conclusion}
Thus, we demonstrated the SPP propagation on the Ag film surface with a record length of propagation $L_{SPP}$ = 194 $\mu$m that is two times longer than previously reported. The measured SPP propagation length corresponds to the calculated data, in which the optical constants of the silver film are taken from [25]. The carried out measurements convincingly show the extremely high quality of the silver films created by the SCULL process [16]. The demonstrated possibility of the enhanced SPP propagation opens up new horizons in the development elements of nanoplasmonics, based on the use of silver films having low losses.

\section*{Acknowledgments}
Silver film samples were made at the  BMSTU Nanofabrication Facility (Functional \\ Micro/Nanosystems, FMNS REC, ID 74300). This work was performed using equipment of MIPT Shared Facilities Center and with financial support from the Ministry of Education and Science of the Russian Federation (Grant No. RFMEFI59417X0014)

\end{document}